\documentclass[reprint,showkeys]{revtex4-2}
\pdfoutput=1
\usepackage{graphicx}
\usepackage{bm}
\usepackage{amsmath}
\usepackage{xcolor}
\usepackage[colorlinks=true,linkcolor=blue,urlcolor=blue,citecolor=blue]{hyperref}

\usepackage{graphicx}
\usepackage{bm}
\usepackage{amsmath}
\usepackage{dcolumn}
\usepackage[colorlinks=true,linkcolor=blue,urlcolor=blue,citecolor=blue]{hyperref}

\newcommand{\rcm}{\mbox{cm$^{-1}$}}

\newcommand{\Est}{E$^1\Sigma^{+}_{\mathrm{u}}$}
\newcommand{\Xst}{X$^1\Sigma^{+}_{\mathrm{g}}$}
\newcommand{\Sst}{$^1\Sigma^{+}_{\mathrm{u}}$}
\newcommand{\Pst}{$^1\Pi_{\mathrm{u}}$}
\newcommand{\norm}[1]{\lvert\lvert #1 \rvert\rvert}

\begin{document}

	\title{The double minimum E(3)$^1\Sigma^{+}_{\mathrm{u}}$ state in Cs$_2$}

	\author{Wlodzimierz Jastrzebski}
	\email{jastr@ifpan.edu.pl}
	\affiliation{Institute of Physics, Polish Academy of Sciences,
		al.~Lotnik\'{o}w~32/46, 02-668~Warsaw, Poland}
	\author{Pawel Kowalczyk}
	\email{Pawel.Kowalczyk@fuw.edu.pl}
	\affiliation{Institute of Experimental Physics, Faculty of Physics,
		University of Warsaw, ul.~Pasteura~5, 02-093~Warszawa, Poland}
	\author{Jacek Szczepkowski}	
	\affiliation{Institute of Physics, Polish Academy of Sciences,
		al.~Lotnik\'{o}w~32/46, 02-668~Warsaw, Poland}
	\author{Asen Pashov}
	\email{pashov@phys.uni-sofia.bg}
	\affiliation{Faculty of Physics, Sofia University, 5~James
	Bourchier blvd., 1164~Sofia, Bulgaria}

	\begin{abstract}
			The double minimum \Est\ state in caesium dimer was investigated by analysing spectra of the \Est\ $\leftarrow$ \Xst\ band system, simplified by polarisation labelling. A total of 6727 rotationally resolved transitions to levels situated in the inner well and above the internal barrier were identified and a potential energy curve allowing to reproduce their energies was constructed using the Fourier grid Hamiltonian and inverted perturbation approach methods. The unambiguousness of the potential curve in view of lack of data related to levels located in the outer well is discussed.
	\end{abstract}
	
	\date{\today}
	
	\maketitle

\section{Introduction}

Electronic states with two minima in the potential energy curve occupy a special place in studies of diatomic molecules. They are particularly difficult for experimental characterisation because positions of rovibrational levels are irregular in the region close to the internal barrier. In addition, the outer potential well is usually unavailable for excitation from the ground electronic state which makes absolute numbering of vibrational levels confusing. The double minimum states are also difficult for theoretical treatment as the two (or multiple) minima arise from anticrossings of curves of diabatic states. Therefore the shape of the potentials is very sensitive to relative positions of the diabatic curves and comparison with experimental results provides sensitive checks of the quality of calculations. In recent years the double minimum states have raised additional interest in view of cold physics experiments, since their unusually broad potential curves provide favourable conditions for photoassociation of molecules from ultracold atoms. It may be followed by spontaneous (or stimulated) emission into the lowest rovibrational levels of the ground electronic state thus enabling making both externally and internally ultracold molecules.

In this paper we present experimental study of the double minimum E(3)$^1\Sigma^{+}_{\mathrm{u}}$ state in Cs$_2$ for which only fragmentary data related to the inner potential well had been available \cite{Amiot1,Amiot2}. This state was suggested for photoassociation experiments in ultracold environment by Pichler {\it et al.} already 20 years ago \cite{Pichler} (and actually used for hot photoassociation even earlier by Ban {\it et al.} \cite{Ban}) but surprisingly has eluded further spectroscopic investigation and the proposal by Pichler {\it et al.} has never been realised. In the present experiment, using the polarisation labelling laser spectroscopy technique (PLS), we observed more than 6700 spectral lines corresponding to transitions from the ground electronic state \Xst\ of caesium dimer to the investigated excited state, most of them terminating on levels lying above the inner potential barrier. Aided by theoretical calculations by Spies \cite{Spies} we constructed a potential energy curve of the \Est\ state allowing to reproduce energies of the rovibrational levels involved in the observed transitions with accuracy better than 0.05~\rcm. The position and height of the barrier on the potential curve was determined. As a particular benefit, we provide the measured frequencies of transitions between the rovibrational levels of the ground state and levels of the double minimum state spanning the range up to about 700~\rcm\ above the barrier, which can be useful for planning of the long awaited photoassociation experiments.

\section{Experiment}

The experiment was performed using the PLS technique, in a variant based on the V-type optical-optical double resonance scheme employing two independent laser sources. Since the experimental setup was identical to this used in our recent experiment on Rb$_2$ \cite{Pashov1}, only a brief summary will be given here. The Cs$_2$ molecules were prepared in the central part of a heat pipe oven by heating metallic caesium to about 485~K with a background pressure of 4~Torr helium. The pump and probe laser beams were collinearly superimposed in the molecular vapour zone. A home built cw single mode external cavity diode laser, fixed on selected transitions in the B\Pst\ $\leftarrow$ \Xst\ band system of Cs$_2$ \cite{Diemer,Nishimiya}, generated the probe beam which labelled known rovibrational levels in the ground electronic state. Its frequency was measured and actively stabilised using a High-Finesse WS-7 wavemeter. The pump beam was delivered by a pulsed, excimer laser pumped dye laser (LightMachinery IPEX-848 and Lumonics HD500, 2~mJ pulse energy, 0.1~\rcm\ linewidth) and its frequency scanned in the spectral range under investigation, i.e. $20200-21300$~\rcm, using Coumarin~480 dye. The pump light induced transitions from the labelled ($v''$, $J''$) levels to levels of the excited \Est\ state. The intensity of the probe beam was monitored through crossed polarisers placed on either side of the oven, so that double resonance signals appeared on a zero background. The residual probe light emerging behind the second polariser provided excitation spectrum of the E$\leftarrow$X band system. This light was detected by a photomultiplier coupled to the boxcar averager (Stanford SR250). By calibrating the pump laser frequency using simultaneously recorded argon and neon optogalvanic spectra and transmission fringes of a Fabry-P\'{e}rot etalon, an absolute accuracy of determination of the transition frequencies in the observed molecular spectra of about 0.05~\rcm\ was achieved. 

\section{Results of measurements and preliminary analysis}

\begin{figure*}[thp]
	\centering
	\includegraphics[width=.95\linewidth]{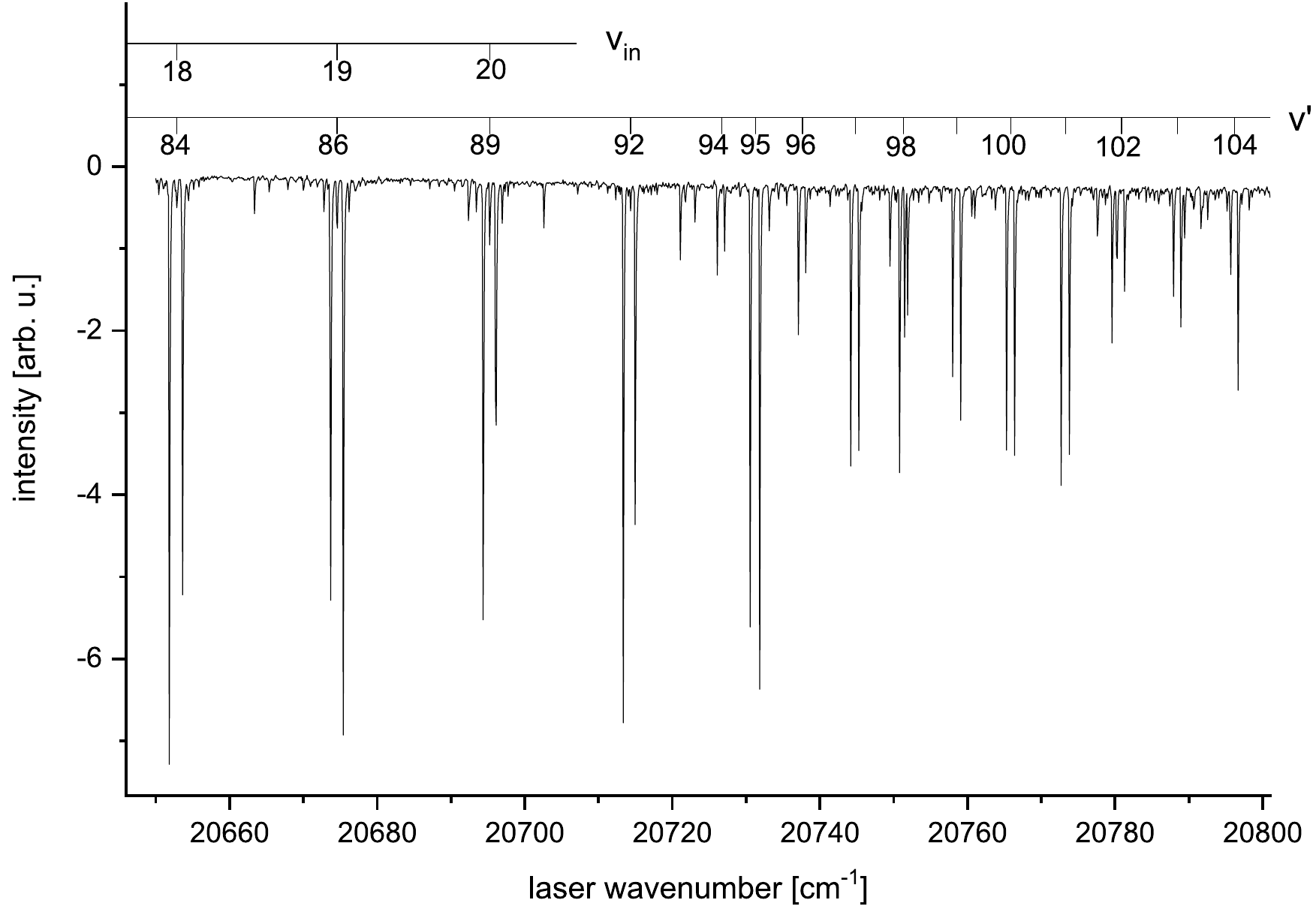}
	\caption {Part of the polarisation spectrum of the E $\leftarrow$ X system displaying vibrational progression originating from the labelled ground state level ($v''=0$, $J''=55$). Transitions to the regions below and above the internal potential barrier can be easily distinguished by a distance of consecutive PR doublets. Quantum numbers $v'$ denote absolute numbering of the vibrational levels, whereas labels $v_{\mathrm{in}}$ correspond to independent numbering only of the levels supported by the inner well. The Dunham coefficients given by Amiot {\it et al.} \cite{Amiot2} allow to reproduce positions of vibrational levels for $v_{\mathrm{in}} \le 19$; for $v_{\mathrm{in}}=20$ discrepancies of 0.3~\rcm\ show that close to the top of the potential barrier the presence of the outer well can no longer be neglected.}

	\label{spectrum}
\end{figure*}

\begin{figure*}[thp]
	\centering
	\includegraphics[width=.95\linewidth]{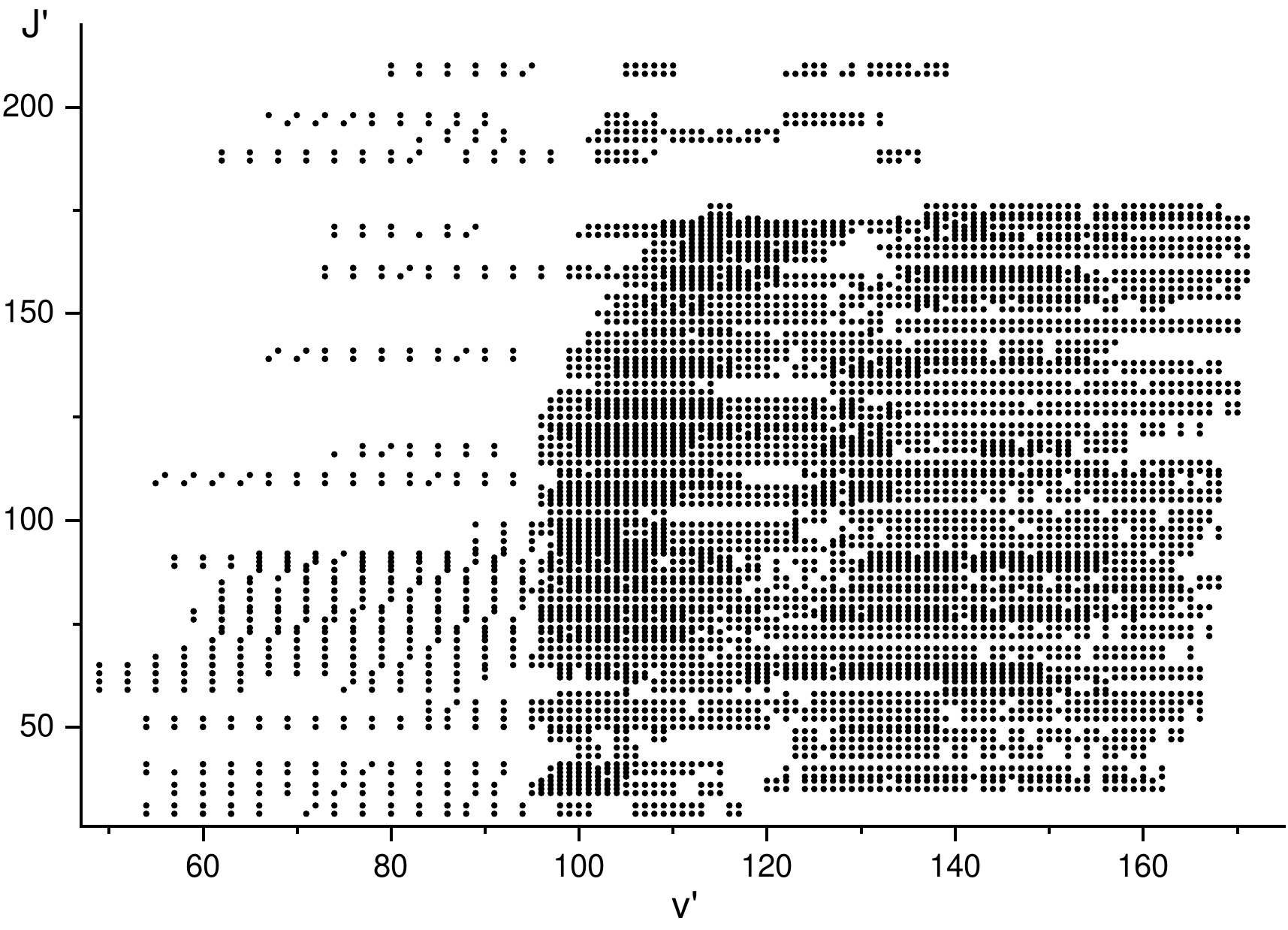}
	\caption {The range of the $v$ and $J$ quantum numbers of the levels in the \Est\ state observed in the present experiment. The numbering of vibrational levels corresponds to the potential energy curve from Table~\ref{table:IPA} (the absolute numbering).}
	\label{data}
\end{figure*}

In the investigated spectral range we have recorded the total of 6727 spectral lines corresponding to transitions between the ground \Xst\ and excited \Est\ state. The transitions started from the purposely labelled levels $v''=0-3$, $J''=30-209$ in the ground state and the $v''=6$, $J''=117 $ level labelled by some spectral coincidence. Due to a negligible overlap between the ground state vibrational wave functions and these corresponding to levels located in the outer well of the double minimum E state, only transitions to the inner potential well and to the region above the barrier were recorded (Figure~\ref{spectrum}). The X state constants of Amiot {\it et al.} \cite{Amiot2} were used to convert the measured transition wave numbers to term values of rovibronic levels in the E state, referred to the minimum of the ground state potential. The experimental errors of Ref. \cite{Amiot2} are about two orders of magnitude smaller than those of our measurements, therefore no additional errors are introduced into the analysis of the \Est\ state. As several upper state levels were reached by transitions originating from more than one ground state level, the number of experimental terms amounts to 5642 (Figure~\ref{data}), smaller than the number of lines. The inner potential well was investigated before in a series of works by Amiot and coworkers and we incorporated to our database 92 term energies related to this well listed in Ref. \cite{Amiot1}. In the present experiment we have concentrated on the previously unexplored levels above the barrier. The assignment of spectral lines in the complex region close to the barrier (Fig.~\ref{spectrum}) was facilitated by the the nature of PLS spectra, consisting of sequences of PR doublets originating from X state levels with known $J''$. In addition, often spectra from ($v'',J''$) and ($v'',J'' + 2$) were recorded, so the energies of some E state terms could have been deduced independently from the two corresponding lines, R$(J'')$ and P$(J''+2)$.

\section{Fitting the potential energy curve of the E\Sst\ state}
\label{Fitting}

Construction of double minimum potential curves of diatomics from the experimental data is a problem which on the scale of complexity lies somewhere between fitting isolated, regular, Morse-like potentials and application of coupled-channels models to mutually perturbed states (see e.g. \cite{Rb2depert}). In the simplest cases one relies on the Born-Oppenheimer approximation, which allows to model the experimental observations with a single-minimum PEC, which eigenvalues are labelled with $v$ and $J$ quantum numbers and the $e$/$f$ symmetry. The vibrational quantum number identifies different eigenvalues for a particular value of $J$. It labels the vibrational wave function and equals to the number of its nodes. At least locally the vibrational levels are nearly equally spaced, so it is possible to identify the consecutive levels from experimental spectra.

In multi-channels problems the eigenvalues cannot be labelled with $v$ anymore, since the vibrational quantum number can be associated only with a single PEC. The multi-channels models are based on several coupled Born-Oppenheimer states, each of them having its own PEC. The solutions of such coupled-channels (CC) problems are superpositions over the initial unperturbed states. We still can assign the eigenvalues of the CC system with ordinal numbers, but they are of limited use as these numbers lose the connection with the vibrational wave functions (and all properties associated with them, like Franck-Condon factors, selection rules etc.). Only in special cases, when the coupling between the basis states is weak, some eigenvalues can be attributed mainly to one of the states and then $v$ can again be used as approximate quantum number to label the vibrational wave functions, since effectively they belong to a single channel PEC. 

Strictly speaking, electronic states with two potential wells  should be treated as a coupled-channels problem because they arise from two coupled diabatic states, the PECs of which cross. In adiabatic approximation one of the resulting PECs has two minima, and both electronic states are still coupled through the off-diagonal matrix element of the operator of kinetic energy of the nuclei \cite{Field}. Depending on the relative strength of the coupling, diabatic or adiabatic approximation may be preferable. In Ref. \cite{Dressler} Dressler proposed a simple criterion:

\[\gamma=\frac{\Delta U_{12}}{\omega_2} \mbox{ ,}\]

\noindent which compares the distance between the adiabatic states at the closest approach, $2\Delta U_{12}$, with the vibrational constant of the upper adiabatic state, $\omega_2$. When $\gamma$ is large ($\gamma>6-7$) the adiabatic picture is more appropriate. For $\gamma\approx 1$ none of the approximations is valid and CC treatment is necessary. In case of the pair of states  E(3)$^1\Sigma^{+}_{\mathrm{u}}$ and 4$^1\Sigma^{+}_{\mathrm{u}}$ in caesium dimer $\gamma > 35$, so one may safely expect accurate results within a single channel adiabatic model. 

There is, however, still a distinct difference between single channel Morse-like and double minimum adiabatic potentials. In the first case it is straightforward to assign the vibrational quantum numbers to the observed energy levels, because their energies $G(v)$ are ordered in a nearly equally spaced sequence. For a double minimum potential $G(v)$ is no longer a smooth, slowly varying function of $v$. This happens in the energy range below and around the internal potential barrier of the double minimum state, where the vibrational levels are located predominantly either in the inner or the outer potential well and follow different vibrational patterns. Adding together positions of levels from both wells results in an apparently haphazard scheme. For a given vibrational level it may be then impossible to predict what is the energy of the next one. Likewise if one fails to observe one or more vibrational levels in the sequence it is not possible to determine what is $v$ of the next observed level. Therefore the analysis of the spectra involving a double minimum state is difficult as the observations provide directly a list of term energies only with well defined $e$/$f$ symmetry and $J$ quantum numbers. There is still another important piece of information contained in the spectra, namely the intensities of the experimental lines. In the \Est $\leftarrow$ \Xst\ system of Cs$_2$ studied in this work, the Franck-Condon factors favour transitions to \Est\ levels for which vibrational wave functions have significant values between approximately 4.3 and 5 \AA\ (compare the ground state PEC from Ref. \cite{Cs2X}). Therefore the levels with vibrational wave functions situated mainly in the outer well could not be observed. 

Facing the problem of constructing potential energy curve of the \Est\ state we could choose between two strategies. The first one may be called an exact-quantum-number strategy. It is assumed that a unique correspondence between the experimental term energies $E^{\mathrm{obs}}_{vJ}$ and the ones calculated from the modelled potential, $E^{\mathrm{calc}}_{vJ}$, can be established. Then the fitting routine tries to minimise the difference between these two sets in a least squares approximation (LSA) sense. This idea lies behind a well known inverse perturbation approach method \cite{Kosman,Vidal,IPAmy,Seto:00, Hadjigeorgiu:00,Samuelis:00}. However, as pointed out before, the problem with double minimum states is that quantum number $v$ cannot be determined from the experimental data and a unique and reliable correspondence cannot be established. As a consequence if one changes the assignment during the fit, the merit function $\chi^2\sim\norm{E^{\mathrm{obs}}_{vJ}-E^{\mathrm{calc}}_{vJ}}$ may become not smooth and thus difficult to minimise. In our previous papers \cite{Pashov1,dmNa2,dmK2,regul,dmNaK} we demonstrated how to overcome this difficulty by assigning provisional $v$ to the experimental data (based on the initial potential curve) and revising this assignment between the consequent iterations. This approach is tedious and requires a lot of attention and manual edition of the current $v$ assignment. For example a small change of the inner well will cause a shift in the inner levels $E_{\mathrm{in}}$, but not in the outer levels $E_{\mathrm{out}}$ and if a particular $E_{\mathrm{in}}$ changes its position with respect to the nearest $E_{\mathrm{out}}$, their vibrational numbers should be interchanged. In practice, such critical pairs of levels should be included to the fit only at the final stage, when the shape of the PEC is nearly fixed by the other experimental data.

In the present work we utilised an alternative approach, namely the nearest-energy strategy, used already by us in some previous CC calculations (\cite{Rb2depert,Szczepkowski:2019}). In this approach the vibrational number $v$ is not used. All levels corresponding to a given $J$ are calculated from the initial potential and then the closest value is searched for every experimental level. Our experience so far shows that this approach works effectively only if the initial guess for the PEC is fairly good. However, the modern {\it ab initio} potentials are already accurate to within few percent. In the present case of the \Est\ state we started with the theoretical PEC provided by Spies \cite{Spies}. Initially we improved only the inner potential well by treating it as independent potential (see \cite{dmNaK}). Here one can assign the so-called inner-well quantum numbers for levels situated well bellow the barrier ($E<20700$~\rcm), cf. Figure~\ref{spectrum}, and fit the PEC by the IPA method \cite{IPAmy}. 

The region above the barrier was also not critical, because $G(v)$ for the vibrational levels is smooth there and relative vibrational numbers can be assigned to them. This part of the potential can be also fitted by the IPA method, after assigning $v$ which corresponds to the nearest levels of the initial potential (as it was done in Ref. \cite{dmNaK}). Alternatively the levels above the barrier can be fitted by the nearest-energy strategy. Here we used a single channel version of the Fourier-Grid-Hamiltonian (FGH \cite{FGH}) code from Ref. \cite{Rb2depert}.

The challenging part of the refinement of the potential was to fit a shape of the barrier influenced mainly by positions of levels situated close to it. We used again the FGH code and the nearest-energy strategy. The fit converged very quickly and in few iterations we finally achieved an agreement with the experiment of 0.043~\rcm. Note that the bottom part of the outer well of \Est\ potential was not fitted and it was fixed to the theoretical shape from Ref. \cite{Spies}.

The grid points defining the potential are listed in Table~\ref{table:IPA}. To interpolate the value of the potential to an arbitrary middle point the natural cubic spline \cite{Boor} should be used, that is the second derivatives at the first and last point should be set to zero. Energies of rovibrational levels up to $v'=171$ and $J'=210$ can be calculated by solving the Schr\"{o}dinger equation in a mesh of 9000 points between 2.8 and 20.0~\AA. 

It should be noted at this juncture, that above the internal barrier where both potential wells merge, the pattern of rovibrational levels regains regularity and their energies can be locally fitted with Dunham-like coefficients (see e.g. \cite{dmK2}), according to

\begin{equation}
	E(v,J)=A_{0,0}+\sum_{m,n}A_{mn}(v'+0.5)^{m}[J'(J'+1)]^{n} \mbox{ .} \label{1}
\end{equation}

\noindent The coefficients shown in Table~\ref{table:Dunhams} allow to reproduce energies of levels in the range $v'=101-171$ and $J'=29-198$ with an rms error of 0.04~\rcm, although they lack direct physical meaning (note for example the negative value of $A_{10}$).

\begin{table}
	
	\centering
	\caption{The rotationless IPA potential energy curve of
		the \Est\ state in Cs$_2$, reproducing the term values of 5631~levels out of a total 5642 measured, with root-mean-square deviation 0.043~\rcm.}
	\label{table:IPA}
	\begin{tabular*}{0.99\linewidth}{@{\extracolsep{\fill}}rrrr}
		&&&\\\hline
		R [\AA] & U [cm$^{-1}$]& R [\AA] & U [cm$^{-1}$]\\ \hline
		&&& \\
		2.80  &  32390.9709    &     8.29 &  20065.3431  \\
		3.00  &  30537.7950    &     8.69 &  19922.8072  \\
		3.20  &  28707.4836    &     9.10 &  19837.9853  \\
		3.41  &  26924.2680    &     9.51 &  19821.0373  \\
		3.61  &  25250.0294    &     9.91 &  19869.9035  \\
		3.81  &  23798.8372    &    10.32 &  19977.8954  \\
		4.02  &  22672.7896    &    10.73 &  20131.9738  \\
		4.17  &  22161.1000    &    11.13 &  20322.3314  \\
		4.36  &  21421.3408    &    11.54 &  20571.3961  \\
		4.52  &  20989.4661    &    11.95 &  20821.1705  \\
		4.62  &  20785.5202    &    12.35 &  21060.5912  \\
		4.75  &  20575.3838    &    12.76 &  21300.7431  \\
		4.86  &  20433.8753    &    13.17 &  21522.9863  \\
		5.01  &  20302.1363    &    13.57 &  21715.9471  \\
		5.16  &  20224.2669    &    14.74 &  21945.0000  \\
		5.33  &  20195.4541    &    16.38 &  22061.2000  \\
		5.47  &  20208.7741    &    19.07 &  22137.6661  \\
		5.65  &  20263.0339    &    22.33 &  22166.6082  \\
		5.83  &  20346.9858    &    25.58 &  22176.8709  \\
		6.02  &  20454.9418    &    27.21 &  22179.3851  \\
		6.20  &  20560.1393    &    30.46 &  22182.1494  \\
		6.44  &  20683.3534    &    33.72 &  22183.4496  \\
		6.73  &  20762.4727    &    36.97 &  22184.1081  \\
		7.00  &  20748.1187    &    40.23 &  22184.4624  \\
		7.27  &  20667.3247    &    43.48 &  22184.6629  \\
		7.47  &  20517.2890    &    46.74 &  22184.7814  \\
		7.88  &  20267.3404    &    50.00 &  22184.8540  \\
				
		&&&\\\hline
			
	\end{tabular*}
\end{table}

\begin{table}
	\centering
	\caption{Dunham-type coefficients representing term values for rovibrational levels of the \Est\ state in Cs$_2$ located above the internal potential barrier. All values are given in \rcm. The coefficients are meaningful only in the range $101 \leq  v{'} \leq 171$ and $29 \leq
		 J{'} \leq 198$.} \label{table:Dunhams} \vspace*{0.5cm}
	\begin{tabular}{cccccc}
		\hline
		& Coefficient & Value (cm$^{-1}$) & Coefficient & Value (cm$^{-1}$) & \\
		\hline \
		& $A_{00}$ &  25759.435                & $A_{01}$ &  0.007756               & \\
		& $A_{10}$ &  -227.183                & $A_{11}$ &  -5.621$\times10^{-5}$ & \\
		& $A_{20}$ &  4.0341                & $A_{21}$ &  3.582$\times10^{-7}$  & \\
		& $A_{30}$ &  -0.03729              & $A_{31}$ & -8.52$\times10^{-10}$   & \\
		& $A_{40}$ &  0.0001956             & $A_{02}$ &  7.52$\times10^{-10}$  & \\
		& $A_{50}$ &  -5.49$\times10^{-7}$  & $A_{12}$ &  -1.62$\times10^{-11}$  & \\		
		& $A_{60}$ &  6.421$\times10^{-10}$  & $A_{03}$ &  5.4$\times10^{-15}$  & \\

		\hline
	\end{tabular}
\end{table}

\section{Discussion}

Very seldom it is possible to determine empirically the whole double well potential curve \cite{dmNa2}. Usually experimental data in the outer well are missing and one has to rely on the shape of the potential predicted by the theory. This raises a question, how realistic is the fitted PEC? Is it only the shape of the outer well which cannot be determined or does it concern the whole PEC?

Concerning the inner well, in our opinion the fit is as reliable as for any single well potential. Sufficiently below the barrier, the vibrational wave functions have significantly non-zero values either only in the inner well or only in the outer well, so a change of the shape of the outer well does not affect the inner well levels. The same is valid for the outer well, but as usually there are no experimental observations of levels supported by this well, its shape cannot be determined in a unique way. The energies of levels above the potential barrier understandably depend on the whole PEC. Since the outer well is undefined, one may expect the same for the upper part of the potential, with a correlation of this part with the shape of the outer well. 

\begin{figure*}[thp]
  \centering
 \includegraphics[width=.95\linewidth]{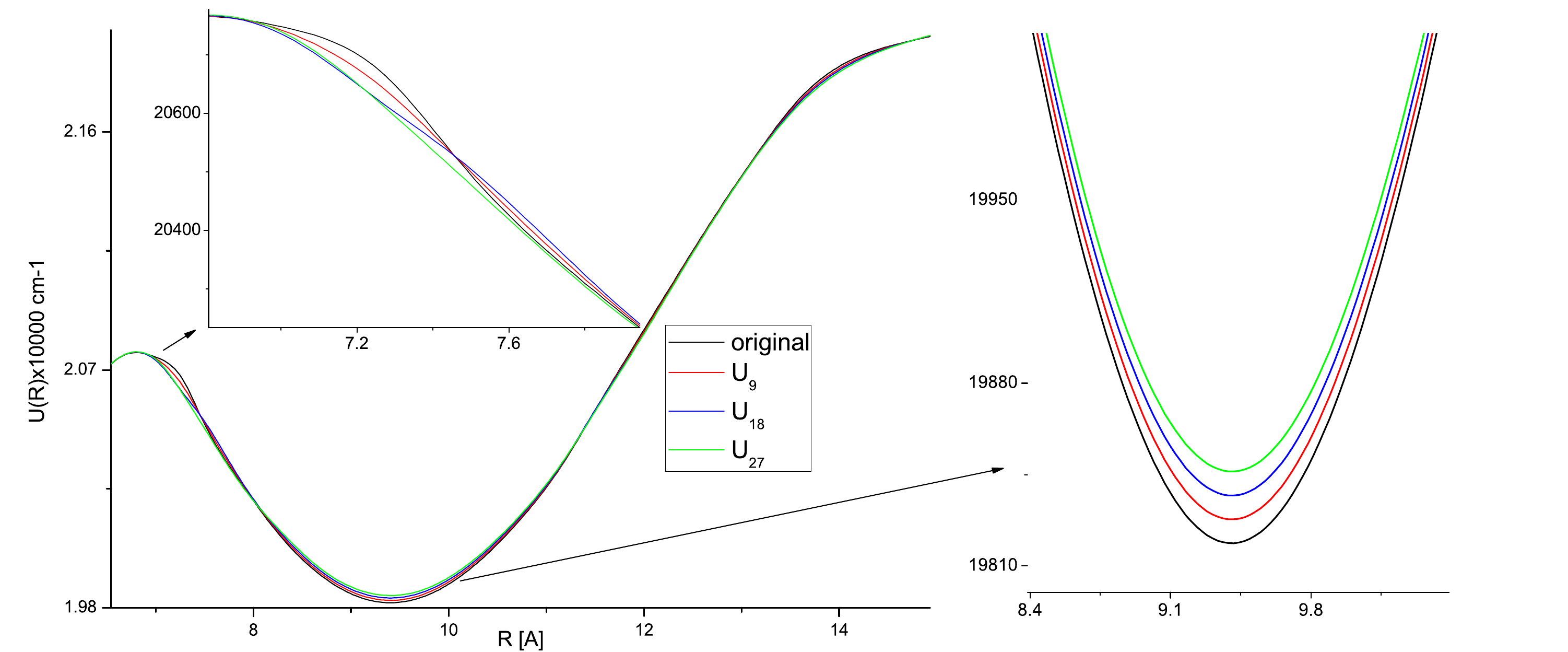}
   \caption {Comparison of parts of four potentials providing similar quality of fit to the experimental data and having different shape of the outer potential well.}
  \label{Pluses}
 \end{figure*}

To shed some light on this problem, we performed a numerical experiment and shifted the bottom of the outer well, adding to the potential of Table~\ref{table:IPA} between 7.27~\AA\ and 11.54~\AA\ ``distortions'' of a functional form $\delta V(R)=a \cdot (11.54 $[\AA] $- R)  \cdot (R - 7.27 $[\AA]), the parameter $a$ being 2~\rcm\AA$^{-2}$, 4~\rcm\AA$^{-2}$ or 6~\rcm\AA$^{-2}$. This resulted in upward shifts of the minimum of the outer well approximately by 9, 18 and 27~\rcm\, respectively (see Figure~\ref{Pluses}), therefore we denote the distorted potentials as $U_{9}$, $U_{18}$ and $U_{27}$. In case of each potential before the fit the rms deviation to our experimental data was about 2~\rcm, so the shift of the potential had significant influence on the experimental energies. After the shift the points around the bottom of the outer well were fixed again (as done previously in section \ref{Fitting}) and the rest of the potential refitted with the nearest-energy strategy. In each case the rms was reduced to below 0.1~\rcm\ within few iterations and to about 0.045~\rcm\ after some points in the outer well were released. As a result we obtained three PECs (shown in Figure~\ref{Pluses}) with slightly different outer wells, different vibrational numbering and the same quality regarding the experimental data.

\begin{figure*}[thp]
  \centering
  \includegraphics[width=.95\linewidth]{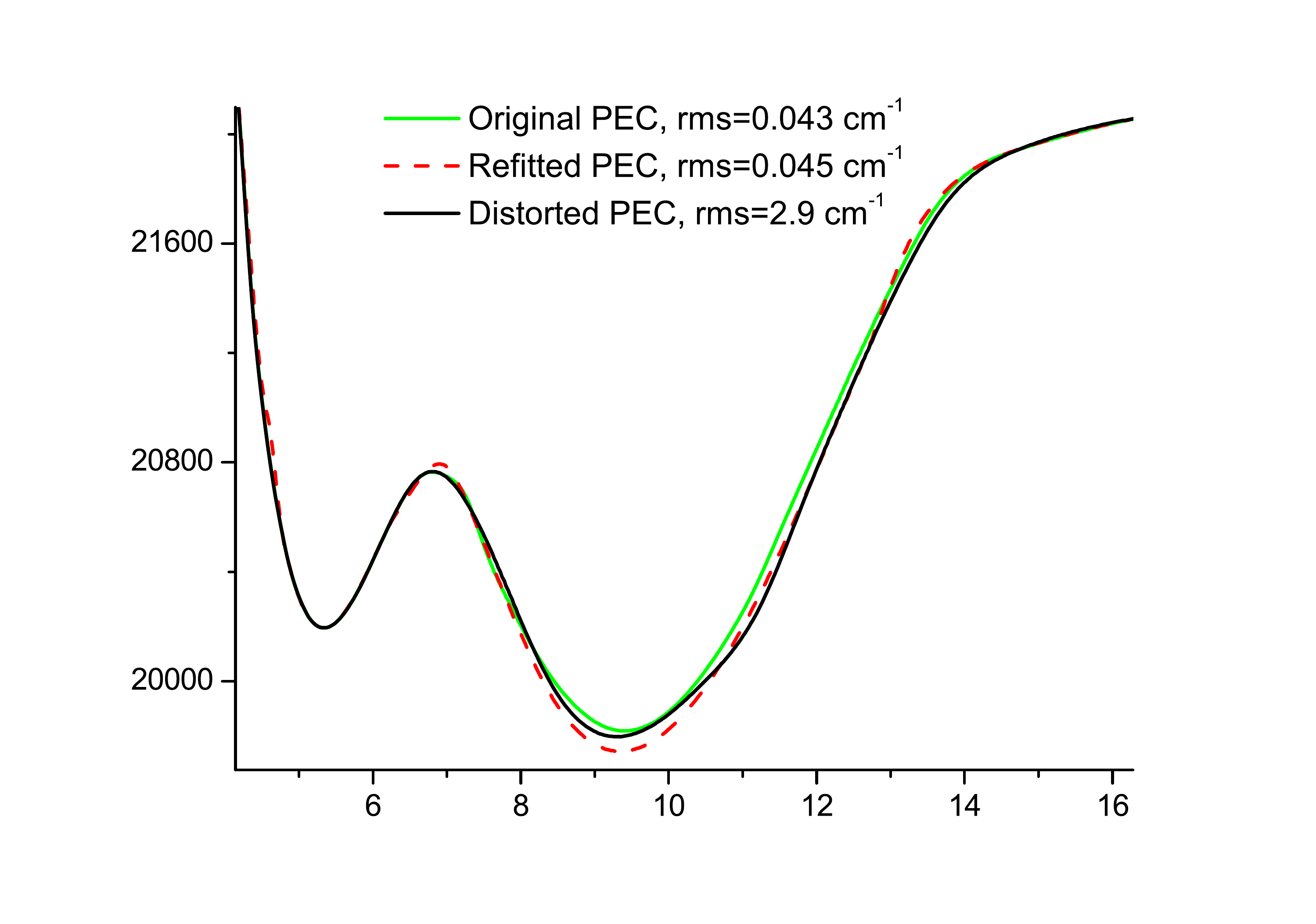}
   \caption {The best potential (green), distorted (black) and the refitted $U_p$ (red, dashed line).} 
  \label{Parabola}
 \end{figure*}

In a more brutal numerical experiment, not only the shape of the outer well but also the rest of the potential were changed more severely (see Figure~\ref{Parabola}). Initially the rms deviation from the experimental data was about 3~\rcm. The distorted potential was fitted only with the closest energy strategy. The fit was much more difficult than in the previous ones. Initially only the inner well was fitted and then fixed. Then the whole potential was subjected to the fit. Few times it was necessary to correct unphysical solutions (as ripples on the potential curve) by manually shifting some points defining the potential. The final result (denoted by us as $U_p$ and providing rms = 0.045~\rcm) is shown in Figure~\ref{Parabola}. One may see that the distorted potential (black curve) does not display a ``regular'' shape. The refitted potential $U_p$ (red dashed curve) is different from the potential of Table~\ref{table:IPA} in the outer well, but also at larger internuclear distances. One may speculate about the meaning of a difference in shape of both potentials around 11~\AA, however the main conclusion of the numerical experiments is that without experimental data related to levels in the outer well not only a shape of this well, but also of parts above the barrier cannot be determined reliably.

Note that even small changes of the potential may influence the numbering of vibrational levels. For example the lowest level in the inner well corresponds in the potential from Table~\ref{table:IPA} to the absolute vibrational quantum number $v=28$ and the same holds for $U_9$. For $U_{18}$ and $U_{27}$ this level is labelled as $v=27$, whereas in the $U_p$ potential as $v=31$. The level with $v=180$ in the potential from Table~\ref{table:IPA} has the same vibrational number in all $U_9$, $U_{18}$ and $U_{27}$, but in $U_{p}$ it corresponds to $v=183$.  The situation might be different if stable isotopologues of Cs existed, because even fragmentary observations of another isotopologue allow to determine the absolute vibrational numbering unambiguously \cite{iso}. As an example, in a similar study of the double minimum 2\Sst\ state in K$_2$ \cite{dmK2} we demonstrated that by combining data from $^{39}$K$_2$ and $^{39}$K$^{41}$K it was possible to establish absolute vibrational numbering of levels. The actual shape of the outer potential well remained uncertain but the number of vibrational levels was fixed by the multi-isotope analysis.

\section{Conclusion}

Basing on a vast number of observed spectral transitions, we have constructed a potential energy curve of the double minimum E(3)$^1\Sigma^{+}_{\mathrm{u}}$ state in Cs$_2$, which allows to accurately reproduce positions of rovibrational levels in the inner potential well and in a broad energy range above the internal potential barrier. A shape of the outer well remains uncertain as its determination would require observation of levels supported by it, unaccessible by the experimental technique of the present study. Nevertheless, this problem does not affect precise description of levels in the regions covered by our experiment.

Suplementary data associated with this article (wavenumbers of the observed spectral lines, energies of the E state rovibrational levels, the IPA potential energy curve of Table~\ref{table:IPA} and the Dunham coefficients of Table~\ref{table:Dunhams}) can be found in its online version as well as at the address \url{http://dimer.ifpan.edu.pl}.

\section{Acknowledgements}

This work was partially funded by the National Science Centre of Poland (Grant No. 2021/43/B/ST4/03326) and by the Bulgarian National Science Fund through Grant K$\Pi$-06-H68/30.11.2022. We also thank for support from the Excellence Initiative -- Research University Program of the University of Warsaw.


\begin{thebibliography}{25}

\bibitem{Amiot1}
C.~Amiot, C.~Cr\'{e}pin, and J.~Verg\`{e}s, Ar$^+$ Laser-Induced Fluorescence Spectra of Cs$_2$: The \Est\ and (1)$^1\Pi_{\mathrm{g}}$ Electronic States, J. Mol. Spectrosc. 107, 28--47 (1984).
	
\bibitem{Amiot2}
C.~Amiot, W.~Demtr\"{o}der, and C.~R.~Vidal, High resolution Fourier spectroscopy and laser spectroscopy of Cs$_2$: The 2$^1\Sigma^{+}_{\mathrm{g}}$, (C) 2\Pst\, (D) 2\Sst\, 3$^1\Sigma^{+}_{\mathrm{g}}$, and (E) 3\Sst\ electronic states, J. Chem. Phys. 88, 5265--5281 (1988).

\bibitem{Pichler}
M.~Pichler, W.~C.~Stwalley, R.~Beuc, and G.~Pichler, Formation of ultracold Cs$_2$ molecules through the double-minimum Cs$_2$ 3\Sst\ state, Phys. Rev. A69, 013403 (2004).

\bibitem{Ban}
T.~Ban, S.~Ter-Avetisyan, R.~Beuc, H.~Skenderovi\'{c}, and G.~Pichler, Photoassociation of cesium atoms into the double minimum Cs$_2$ 3\Sst\ state, Chem. Phys. Lett. 313, 110--114 (1999).

\bibitem{Spies}
N.~Spies, \emph{Theoretische Untersuchung von elektronisch angeregten Zust\"{a}nden der Molek\"{u}le Li$_2$ und Cs$_2$}, PhD Thesis, Universit\"{a}t Kaiserslautern, 1990.

\bibitem{Pashov1}
A.~Pashov, P.~Kowalczyk, and W.~Jastrzebski, Double-minimum 3\Sst\ state in Rb$_2$: Spectroscopic study and possible applications for cold-physics experiments, Phys. Rev. A 100, 012507 (2019).

\bibitem{Diemer}
U.~Diemer, R.~Duchowicz, M.~Ertel, E.~Mehdizadeh, and W.~Demtr\"{o}der, Doppler-free polarization spectroscopy of the B\Pst\ state of Cs$_2$, Chem. Phys. Lett. 164, 419--426 (1989).

\bibitem{Nishimiya}
 N.~Nishimiya, Y.~Yasuda, T.~Yukiya, and M.~Suzuki, Sub-Doppler polarization spectroscopy of the B\Pst\ $\leftarrow$ \Xst\ system for Cs$_2$ with a titanium sapphire ring laser, J. Mol. Spectrosc. 255, 194--198 (2009).

\bibitem{Rb2depert} 
A.~Pashov, P.~Kowalczyk, A.~Grochola, J.~Szczepkowski, and W.~Jastrzebski, Coupled-channels analysis of the (5\Sst, 5\Pst, $5^3\Pi_u$, $2^3\Delta_u$) complex of electronic states in rubidium dimer, J. Quant. Spectrosc. Radiat. Transfer, 22, 225--232 (2018).

\bibitem{Field}
H.~Lefebvre-Brion and R.~W.~Field, \emph{The Spectra and Dynamics of Diatomic Molecules}, Elsevier, Amsterdam, 2004.

\bibitem{Dressler}
K.~Dressler, in \emph{Photophysics and Photochemistry above 6 eV}, ed. F. Lahmani, p. 327, Elsevier, Amsterdam, 1985.

\bibitem{Cs2X}
C.~Amiot and O.~Dulieu, The Cs$_2$ ground electronic state by Fourier transform spectroscopy: Dispersion coefficients, J. Chem. Phys. 117, 5155--5164 (2002).

\bibitem{Kosman}
W.~Kosman and J.~Hinze, Inverse perturbation analysis: Improving the accuracy of potential energy curves, J. Mol. Spectrosc. 56, 93--103 (1975).

\bibitem{Vidal}
C.~Vidal and H.~Scheingraber, Determination of diatomic molecular constants using an inverted perturbation approach: Application to the A{$^1\Sigma^{+}_{\mathrm{u}}$}--X{$^1\Sigma^{+}_{\mathrm{g}}$} system of Mg$_2$, J. Mol. Spectrosc. 65, 46--64 (1977).

\bibitem{IPAmy}
A.~Pashov, W.~Jastrz\c{e}bski, and P.~Kowalczyk, Construction of potential curves for diatomic molecular states by the IPA method, Comp. Phys. Commun. 128, 622--634 (2000).

\bibitem{Seto:00}
J.~Y.~Seto, R.~J.~Le~Roy, J.~{Verg\`{e}s}, and C.~Amiot, Direct potential fit analysis of the X$^1\Sigma^{+}_{\mathrm{g}}$   state of Rb$_2$. Nothing else will do!, J. Chem. Phys. 113, 3067--3076 (2000).

\bibitem{Hadjigeorgiu:00}
P.~G.~Hajigeorgiou and R.~J.~Le~Roy, A ``Modified Lennard-Jones oscillator'' model for diatom potential functions, J. Chem. Phys. 112, 3949--3957 (2000).

\bibitem{Samuelis:00}
C.~Samuelis, E.~Tiesinga, T.~Laue, M.~Elbs, H.~{Kn\"{o}ckel}, and E.~Tiemann,
Cold atomic collisions studied by molecular spectroscopy, Phys. Rev. A63, 012710 (2000).

\bibitem{dmNa2}
A.~Pashov, W.~Jastrz\c{e}bski, W.~Ja\'{s}niecki, V.~Bednarska, and P.~Kowalczyk, Accurate potential curve of the double minimum 2\Sst\ state of Na$_2$, J. Mol. Spectrosc. 203, 264--267 (2000).

\bibitem{dmK2}
W.~Jastrz\c{e}bski, W.~Ja\'{s}niecki, P.~Kowalczyk, R.~Nadyak, and A.~Pashov, Spectroscopic investigation of the double minimum 2\Sst\ state of potassium dimer, Phys. Rev. A62, 042509 (2000).

\bibitem{regul}
A.~Grochola, P.~Kowalczyk, W.~Jastrzebski, and A.~Pashov, A regularized inverted perturbation approach method: potential energy curve of the 4\Sst\ state in Na$_2$, J. Chem. Phys. 121, 5754--5760 (2004).

\bibitem{dmNaK}
A.~Pashov, W.~Jastrz\c{e}bski, and P.~Kowalczyk, Improved description of the double minimum $6^1\Sigma^+$ state of NaK by an IPA potential energy curve, J. Phys. B: At. Mol. Opt. Phys. 33, L611--L614 (2000).

\bibitem{Szczepkowski:2019}
J.~Szczepkowski, A.~Grochola, P.~Kowalczyk, W.~Jastrzebski, E.~A.~Pazyuk, A.~V.~Stolyarov, and A.~Pashov, The spin-orbit coupling of the {6$^1\Sigma^+$ and 4$^3\Pi$} states in KCs: observation and deperturbation, J. Quant. Spectrosc. Radiat. Transfer, 239, 106650 (2019).

\bibitem{FGH}
C.~C.~Marston and G.~G.~Balint-Kurti, The Fourier grid Hamiltonian method for bound state eigenvalues and eigenfunctions, J. Chem. Phys. 91, 3571--3576 (1989).

\bibitem{Boor}
C.~De Boor, \emph{A Practical Guide to Splines}, Springer, Berlin, 1978.

\bibitem{iso}
A.~Pashov, P.~Kowalczyk, and W.~Jastrzebski, Absolute vibrational numbering from isotope shifts in fragmentary spectroscopic data, J. Mol. Spectrosc. 347, 48--55 (2018).

\end{thebibliography}
\end{document}